\documentclass{JHEP3}
\usepackage{amsmath}
\usepackage[sort&compress,numbers]{natbib}

\def\eps{\epsilon}

\def\Poles{{\cal P}oles}

\def\e{\epsilon}

\preprint{
ZU--TH 24/14}

\title{The $Hb\bar{b}$ form factor to three loops in QCD}

\author{Thomas Gehrmann and Dominik Kara
	\\
Physik-Institut, Universit\"at Z\"urich, Winterthurerstrasse 190, 8057 Z\"urich, 
Switzerland\\
E-mail: \email{thomas.gehrmann@uzh.ch, dkara@physik.uzh.ch}
}

\abstract{We compute the three-loop QCD corrections to the vertex function 
for the Yukawa coupling of a Higgs boson to a pair of bottom quarks in the limit of vanishing quark 
masses. This QCD form factor is a crucial ingredient to third-order QCD corrections for the 
production of Higgs bosons in bottom quark fusion, and for the fully differential decay rate of 
Higgs bosons to bottom quarks. The  infrared pole structure of the 
form factors agrees with the prediction from infrared 
factorization in QCD.}

\begin{document}

\section{Introduction}
With the discovery of the Higgs boson at  the CERN LHC~\cite{higgsLHC}, the full 
 Standard Model spectrum of matter particles and force carriers has been 
 established successfully. To fully validate the mechanism of 
 electroweak symmetry breaking, and to uncover potential deviations from its 
 Standard Model realization, it is imperative to study the production mechanisms and 
 decay channels of the Higgs boson to high precision. 
The interpretation of increasingly accurate experimental  data from the 
upcoming run periods at the LHC demands equally precise theoretical predictions, 
 requiring the inclusion of higher orders in the perturbative expansion for production and 
 decay processes. 
 
Currently, fully differential results are known for Higgs boson production in 
gluon fusion~\cite{ggHnnlo}, bottom quark annihilation~\cite{bbHnnlo} and associated 
production with vector bosons~\cite{vhnnlo} 
to next-to-next-to-leading order (NNLO) in QCD. Vector boson 
fusion~\cite{vbfnlo} and  associated production with top quarks~\cite{ttHnlo} are known to 
next-to-leading order (NLO). The inclusive decay rates of the Higgs boson have been derived to 
fourth order in QCD for the decay mode to hadrons~\cite{Hhad} and to bottom 
quarks~\cite{Hbbar}. To study the dominant decay mode to 
bottom quarks, especially the associated production with vector
bosons is of relevance, and a fully differential description of production and decay is 
demanded. The decay distributions to NNLO have been derived in Ref.~\cite{lazopoulos}, and 
a combined description with the associated production at NNLO was obtained 
recently~\cite{grazzininew}.

In the Standard Model, the dominant Higgs boson production process is gluon fusion, while 
bottom quark annihilation contributes to 
the total production only at the per-cent level. In extensions of the Standard Model with an 
enlarged spectrum in the Higgs sector, the coupling of some of the Higgs bosons to bottom quarks 
can be enhanced, such that bottom quark annihilation could become their dominant 
production process. Bottom quark annihilation is moreover of conceptual interest since it allows 
to study different prescriptions for the treatment of bottom-quark induced processes at hadron 
colliders. In the fixed flavor number scheme (FFNS), bottom quarks are produced 
only from gluon splitting, introducing potentially large logarithmic corrections at each order. 
These initial-state splittings are resummed into bottom quark parton distributions in 
the variable flavor number scheme (VFNS). To the same order in the strong coupling constant,
the leading order process in the FFNS corresponds to NNLO in the VFNS. 
Higgs production from bottom quark annihilation is known to NLO in the FFNS~\cite{ffns} and 
to NNLO in the VFNS~\cite{bbHnnlo,Harlander:2003,Harlander:2014}. 
Using the NLO calculation of Higgs-plus-jet production in bottom quark 
annihilation~\cite{Harlander:2010cz}, the Higgs production with a jet veto was also 
derived to NNLO~\cite{Harlander:2011fx}.
Most calculations are 
carried out in the limit of vanishing 
bottom quark mass, which is justified by the large mass hierarchy between the bottom quark and 
the Higgs boson. 

The calculation of perturbative higher order QCD corrections requires the derivation of 
virtual loop corrections to the relevant matrix elements. In the case of Higgs production and 
decay involving bottom quarks, the form factor describing the Yukawa coupling of the 
Higgs boson to bottom quarks is the crucial ingredient. Corrections up to two loops were derived 
for this form factor for massless bottom quarks~\cite{bbHnnlo,Harlander:2003,Ravindran:2006} 
and also including the full 
mass dependence~\cite{Bernreuther:2005}. The pole structure of the massless form factor at three 
loops can be predicted~\cite{Ravindran:2006} from factorization properties of QCD 
amplitudes~\cite{catani,sterman,MMV1,Becher:2009cu,Gardi:2009qi}.
Two-loop corrections to the Higgs decay 
amplitude describing the decay to a pair of bottom quarks and a gluon were 
also derived~\cite{Ahmed:2014} in massless QCD. 
It is the aim of the present paper to derive the three-loop 
QCD corrections to the $Hb\bar b$ form factor, working in the limit of vanishing $b$-quark mass. 
This form factor enters the N$^3$LO corrections to the  Higgs production cross section from 
bottom quark annihilation and the differential description of Higgs decays to bottom quarks at 
this order. Both types of applications require a substantial extension of current technical methods 
in order to perform calculations of collider observables to N$^3$LO. First steps in this direction 
have been taken recently~\cite{n3lomethods}, cumulating in the calculation of the N$^3$LO 
threshold contribution to Higgs production in gluon fusion~\cite{claudehiggs}. Exploiting 
universal QCD factorization properties at threshold \cite{ravindran1,grazziniRV}, the 
result of Ref.~\cite{claudehiggs} can be combined with the 
form factor derived here to obtain the N$^3$LO threshold 
contribution to Higgs production in bottom quark annihilation. 

This paper is structured as follows: in Section~\ref{sec:renorm} we define the $Hb\bar b$ form factor,
discuss its renormalization and 
summarize results at one and two loops. The calculation of the three-loop form factor 
proceeds along the lines of the calculations of the three-loop QCD corrections to 
the vector and scalar form factors~\cite{BCSSS,Gehrmann:2010} and is 
described in Section~\ref{sec:red}. The results are presented in Section~\ref{sec:ff} and the infrared pole structure is analyzed in Section~\ref{sec:ir}.
We conclude with an outlook in Section~\ref{sec:conc}. 

\section{$Hb\bar{b}$ form factor in perturbative QCD}
\label{sec:renorm}

In general, form factors are scalar functions which couple an external off-shell current with four-momentum $q^2 = s_{12}$ to a pair of partons
with on-shell momenta $p_1$ and $p_2$. They are computed by contracting the respective basic vertex functions with projectors.
In the $Hb\bar{b}$ case, the unrenormalized form factor $\cal F$ is obtained from a scalar vertex function $\Gamma$ according to
\begin{equation}
{\cal F}  = -\frac{1}{2 q^2}\, {\mathrm Tr} \left( p_1 \!\!\!\! / \, p_2 \!\!\!\! / \, \Gamma \right)\, , \label{eq:projq}
\end{equation}
where $p_i \!\!\!\! / \, = p_{i,\mu} \gamma^\mu$. Note that the vertex function is evaluated in dimensional regularization with $D=4-2\e$ dimensions. It is described by a single form factor only in the case of massless partons. In fact, we consider a Higgs boson coupling to the bottom quarks via an unrenormalized Yukawa coupling $y^b$,
\begin{equation}
y^b = \frac{m^b}{v} \, ,
\end{equation}
with the Standard Model Higgs vacuum expectation value $v=(\sqrt{2} G_F)^{-1/2}$, Fermi's constant $G_F$ and the bare mass $m^b$ of the bottom quark. However, we treat the bottom mass as independent of the Yukawa coupling and suppose it to be massless in the calculation of matrix elements. This is justified by the fact that the Higgs boson is much heavier than the bottom quark.

Evaluating the Feynman diagrams contributing to the vertex function in perturbative QCD at a given loop order yields the unrenormalized form factor as an expansion in powers of the coupling constant. With the mass parameter $\mu_0^2$, which is introduced in dimensional regularization to maintain a dimensionless coupling in the bare Lagrangian density, and the definition
\begin{equation}
S_{\eps} = e^{-\eps \gamma} (4\pi)^{\eps}, \qquad \qquad {\rm with~the~Euler~constant~} \gamma =  0.5772\ldots,
\end{equation}
this expansion can be written as
\begin{eqnarray}
{\cal F} (\alpha_s^b, s_{12}) &=& y^b\left(1 + \sum_{n=1}^{\infty} \left( \frac{\alpha_s^b}{4\pi}\right)^n \left(\frac{-s_{12}}{\mu_0^2}\right)^{-n\eps}
S_{\eps}^n \,{\cal F}_n\right) \, .
\end{eqnarray}
Each power of the coupling constant corresponds to a virtual loop, i.e. Eq.~\eqref{eq:projq} is normalized in such a way that the tree-level form factor is equal to unity.\\
The ultraviolet (UV) renormalization of the form factor requires two ingredients: First, the bare coupling $\alpha_s^{b}$ is replaced with the renormalized coupling $\alpha_s\equiv \alpha_s(\mu^2)$, which is evaluated at the renormalization scale $\mu^2$:
\begin{equation}
\alpha_s^b \mu_0^{2\epsilon} = Z_{\alpha_s} \mu^{2\epsilon}  \alpha_s(\mu^2) \, .
\end{equation}
For the sake of clarity, we set $\mu^2 = |s_{12}|$ throughout so that the renormalization constant of the strong coupling in the $\overline{{\rm MS}}$ scheme \cite{msbar} reads 
\begin{eqnarray}
Z_{\alpha_s} &=& S_{\eps}^{-1}\Bigg[  
1- \frac{\beta_0}{\e}\left(\frac{\alpha_s}{4\pi}\right) 
+\left(\frac{\beta_0^2}{\e^2}-\frac{\beta_1}{2\e}\right)
\left(\frac{\alpha_s}{4\pi}\right)^2\nonumber \\
&& \hspace{1cm}
-\left(\frac{\beta_0^3}{\e^3}-\frac{7}{6}\frac{\beta_1\beta_0}{\e^2}+\frac{1}{3}\frac{\beta_2}{\e}\right)\left(\frac{\alpha_s}{4\pi}\right)^3+{\cal O}(\alpha_s^4) \Bigg] \, .
\end{eqnarray}
$\beta_0$, $\beta_1$ and $\beta_2$ are the first three coefficients of the QCD beta function expanded in powers of the coupling constant:
\begin{equation}
\beta \left(\alpha_s\right) \equiv \frac{1}{4\pi} \frac{\mathrm d \, \alpha_s}{\mathrm d \, \mathrm{ln} \, \mu^2} = - \beta_0 \left(\frac{\alpha_s}{4\pi}\right)^2 - \beta_1 \left(\frac{\alpha_s}{4\pi}\right)^3 - \beta_2 \left(\frac{\alpha_s}{4\pi}\right)^4 + {\cal O}(\alpha_s^5) \, .
\end{equation}
They are given by \cite{beta0,beta1,beta2}
\begin{eqnarray}
\beta_0 &=& 
\frac{11 C_A}{3}-\frac{2 N_F}{3},\\
\beta_1 &=& 
\frac{34 C_A^2}{3}-\frac{10 C_A N_F}{3}-2 C_F N_F,\\
\beta_2 &=& 
\frac{2857 C_A^3}{54}+C_F^2 N_F-\frac{205 C_F C_A N_F}{18}-\frac{1415 C_A^2 N_F}{54}+\frac{11 C_F N_F^2}{9}+\frac{79 C_A N_F^2}{54}
\end{eqnarray}
with the number of quark colors $N$, the number of active quark flavors $N_F$ and the $SU(N)$ Casimirs
\begin{equation}
C_F = \frac{N^2-1}{2 \, N} \quad \text{and} \quad C_A = N \, .
\end{equation}
Second, the renormalization of the Yukawa coupling is carried out by replacing the bare coupling $y^{b}$ with the renormalized coupling $y\equiv y(\mu^2)$ according to
\begin{equation}
y^b = Z_y y(\mu^2) \, .
\end{equation}
$Z_y$ is identical to the quark mass renormalization constant of QCD in the $\overline{{\rm MS}}$ scheme \cite{Harlander:2003}. It can be evaluated to three loops with the help of the relation
\begin{equation}
\gamma_m = -\frac{\mathrm d \, \mathrm{ln} \, Z_y}{\mathrm d \, \mathrm{ln} \, \mu^2} = -\frac{\partial \, \mathrm{ln} \, Z_y}{\partial \, \alpha_s} \frac{\mathrm d \, \alpha_s}{\mathrm d \, \mathrm{ln} \, \mu^2} = - 4\pi \, \frac{\partial \, \mathrm{ln} \, Z_y}{\partial \, \alpha_s} \left[-\e \, \left(\frac{\alpha_s}{4\pi}\right) + \beta\left(\alpha_s\right) \right]
\end{equation}
from the quark mass anomalous dimension
\begin{equation}
\gamma_m \equiv -\gamma_0 \left(\frac{\alpha_s}{4\pi}\right) - \gamma_1 \left(\frac{\alpha_s}{4\pi}\right)^2 - \gamma_2 \left(\frac{\alpha_s}{4\pi}\right)^3 +{\cal O}(\alpha_s^4) \, ,
\end{equation}
which is given in Ref.~\cite{Vermaseren:1997}. This results in
\begin{eqnarray}
Z_y= 
1 &-& \frac{3 C_F}{\e}\left(\frac{\alpha_s}{4\pi}\right) 
\nonumber \\
&+&\Bigg[C_F^2 \left(\frac{9}{2 \e^2}-\frac{3}{4 \e} \right) + C_F C_A \left(\frac{11}{2 \e^2}-\frac{97}{12 \e}\right) + C_F N_F \left(-\frac{1}{\e^2}+\frac{5}{6 \e}\right)\Bigg]
\left(\frac{\alpha_s}{4\pi}\right)^2
\nonumber \\
&+&\Bigg[
C_F^3 \left(-\frac{9}{2 \e^3}+\frac{9}{4 \e^2}-\frac{43}{2 \e} \right) + C_F^2 C_A \left(-\frac{33}{2 \e^3}+\frac{313}{12 \e^2}+\frac{43}{4 \e} \right)
\nonumber \\
&& \quad + C_F C_A^2 \left(-\frac{121}{9 \e^3}+\frac{1679}{54 \e^2}-\frac{11413}{324 \e} \right) + C_F^2 N_F \left(\frac{3}{\e^3}-\frac{29}{6 \e^2}+\frac{1}{\e} \left(\frac{23}{3}-8\zeta_3\right) \right)
\nonumber \\
&& \quad + C_F C_A N_F \left(\frac{44}{9 \e^3}-\frac{242}{27 \e^2}+\frac{1}{\e} \left(\frac{278}{81}+8\zeta_3\right) \right)
\nonumber \\
&& \quad + C_F N_F^2 \left(-\frac{4}{9 \e^3}+\frac{10}{27 \e^2}+\frac{35}{81 \e} \right)
\Bigg]
\left(\frac{\alpha_s}{4\pi}\right)^3
+{\cal O}(\alpha_s^4) \; .
\end{eqnarray}
The renormalized form factor $F$ is defined as follows:
\begin{eqnarray}
{F} (\alpha_s(\mu^2), s_{12},\mu^2= |s_{12}|) &=& y \left(1 + \sum_{n=1}^{\infty} \left( \frac{\alpha_s(\mu^2)}{4\pi}\right)^n   
  \,{F}_n\right).
\end{eqnarray}
In order to derive the $i$-loop contributions $F_i$ to the renormalized form factor from the unrenormalized coefficients ${\cal{F}}_i$, two possible configurations have to be distinguished: The partons can be both either in the initial or in the final state ($s_{12}>0$, time-like) or one parton can be in the initial and one in the final state ($s_{12}<0$, space-like). We will indicate the results for the renormalized form factors in the time-like case, which corresponds to the Higgs decay into bottom quarks or to the $b\bar{b}$ annihilation process into a Higgs boson. In this case, the renormalized form factor acquires imaginary parts from the $\epsilon$-expansion of
\begin{equation}
\Delta(s_{12}) = (-{\rm sgn}(s_{12})-i0)^{-\epsilon} \, .
\end{equation}
Up to three loops, the renormalized coefficients for the $Hb\bar{b}$ form factor are then obtained as
\begin{eqnarray}
F_1   &=& 
 {\cal F}_1 \Delta(s_{12})
-\frac{3 C_F}{\e}   ,  \nonumber \\
F_2  &=& 
 {\cal F}_2 \left(\Delta(s_{12})\right)^2
+ \Bigg[ -\frac{3 C_F}{\e}-\frac{11 C_A}{3 \e}+\frac{2 N_F}{3 \e} \Bigg]
{\cal F}_1  \Delta(s_{12}) \nonumber \\
&& + \Bigg[ C_F^2 \left(\frac{9}{2 \e^2}-\frac{3}{4 \e}\right) + C_F C_A \left(\frac{11}{2 \e^2}-\frac{97}{12 \e}\right) + C_F N_F \left(-\frac{1}{\e^2}+\frac{5}{6 \e}\right) \Bigg] , \nonumber \\
F_3  &=& 
 {\cal F}_3 \left(\Delta(s_{12})\right)^3
+ \Bigg[ -\frac{3 C_F}{\e}-\frac{22 C_A}{3 \e}+\frac{4 N_F}{3 \e} \Bigg] {\cal F}_2  \left(\Delta(s_{12})\right)^2 \nonumber \\
&& + \Bigg[ C_F^2 \left(\frac{9}{2 \e^2}-\frac{3}{4 \e}\right) + C_F C_A \left(\frac{33}{2 \e^2}-\frac{97}{12 \e}\right) + C_A^2 \left(\frac{121}{9 \e^2}-\frac{17}{3 \e}\right) \nonumber \\
&& \quad + C_F N_F \left(-\frac{3}{\e^2}+\frac{11}{6 \e}\right) + C_A N_F \left(-\frac{44}{9 \e^2}+\frac{5}{3 \e}\right) + \frac{4 N_F^2}{9 \e^2} \Bigg] {\cal F}_1\Delta(s_{12}) \nonumber \\
&& + \Bigg[ C_F^3 \left(-\frac{9}{2 \e^3}+\frac{9}{4 \e^2}-\frac{43}{2 \e}\right) + C_F^2 C_A \left(-\frac{33}{2 \e^3}+\frac{313}{12 \e^2}+\frac{43}{4 \e}\right) \nonumber \\
&& \quad + C_F C_A^2 \left(-\frac{121}{9 \e^3}+\frac{1679}{54 \e^2}-\frac{11413}{324 \e}\right) + C_F^2 N_F \left(\frac{3}{\e^3}-\frac{29}{6 \e^2}+\frac{1}{\e} \left(\frac{23}{3}-8\zeta_3\right) \right) \nonumber \\
&& \quad + C_F C_A N_F \left(\frac{44}{9 \e^3}-\frac{242}{27 \e^2}+\frac{1}{\e} \left(\frac{278}{81}+8\zeta_3\right) \right) \nonumber \\
&& \quad + C_F N_F^2 \left(-\frac{4}{9 \e^3}+\frac{10}{27 \e^2}+\frac{35}{81 \e}\right) \Bigg] .
\label{eq:reng}
\end{eqnarray}
The one- and two-loop relations agree with those in Ref.~\cite{Ahmed:2014}.

\subsection{Results at one loop}

We define
\begin{equation}
 S_{R} = \frac{16\pi^2 S_{\Gamma}}{S_{\epsilon}} = \frac{\exp(\epsilon \gamma)}{\Gamma(1-\epsilon)} \, ,
\label{eq:sR}
\end{equation}
where
\begin{equation}
 S_{\Gamma} = \frac{(4\pi)^\e}{16\pi^2\Gamma(1-\e)}
\label{eq:sgamma}
\end{equation}
corresponds to the normalization of the one-loop bubble integral $B_{2,1}$. With this, the unrenormalized one-loop form factor can be written as
\begin{eqnarray}
\label{eq:f1q}
{\cal F}_1/S_R =
&&C_F\,B_{2,1}
\left(\frac{4}{(D-4)}+D\right) \, .
\label{eq:f1g}
\end{eqnarray}
The exact result for the one-loop bubble integral is indicated in Ref.~\cite{GHM} under the name $A_{2,\mathrm{LO}}$, i.e. Eq.~\eqref{eq:f1g} can be understood as an all-order expression. The $\e$-expansion of $B_{2,1}$ can be found in Appendix~A of Ref.~\cite{Gehrmann:2010}. Inserting this expansion and keeping terms through to ${\cal O}(\eps^4)$, we obtain
{\allowdisplaybreaks
\begin{eqnarray}
{\cal F}_1 = C_F\Biggl[
&&-\frac{2}{\eps^2}
+\left(\zeta_2-2\right)
+\eps\left(\frac{14\zeta_3}{3}-4\right)
+\eps^2\left(\frac{47\zeta_2^2}{20}+\zeta_2-8\right)\nonumber \\
&&-\eps^3\left(\frac{7\zeta_2\zeta_3}{3}-2\zeta_2-\frac{14\zeta_3}{3}-\frac{62\zeta_5}{5}+16\right)\nonumber \\
&&+\eps^4\left(\frac{949\zeta_2^3}{280}+\frac{47\zeta_2^2}{20}-\frac{49\zeta_3^2}{9}+4\zeta_2+\frac{28\zeta_3}{3}-32\right)
%\nonumber \\
%&&+\eps^5\bigg(
%\frac{47\zeta_2^2}{10}
%-\frac{329\zeta_2^2\zeta_3}{60}
%-\frac{7\zeta_2\zeta_3}{3}
%-\frac{31\zeta_2\zeta_5}{5}\nonumber \\
%&&\hspace{1cm}
%+8\zeta_2
%+\frac{56\zeta_3}{3}
%+\frac{62\zeta_5}{5}
%+\frac{254\zeta_7}{7}
%-64\bigg)
\Biggr] \, .
\end{eqnarray}
}

By renormalizing this result as described in Eq.~\eqref{eq:reng}, we find that the one-loop form factor agrees with the $\e$-expansion of Eq.~(3.2) in Ref.~\cite{lazopoulos}.

\subsection{Results at two loops}

Written in terms of the two-loop master integrals specified in Appendix~A of Ref.~\cite{Gehrmann:2010}, the unrenormalized two-loop form factor is given by
{\allowdisplaybreaks
\begin{eqnarray}
{\cal F}_2/S_R^2 = C_F^2 \Biggl[
&&B_{4,2}
\left(D^2+\frac{32}{(D-4)}+\frac{16}{(D-4)^2}+8\right)\nonumber \\
&&-C_{4,1}
\left(\frac{7D^2}{8}-\frac{137D}{16}-\frac{265}{32(2D-7)}-\frac{58}{(D-4)}-\frac{40}{(D-4)^2}-\frac{239}{32}\right)\nonumber \\
&&+ B_{3,1} 
\left(\frac{27D^2}{8}-\frac{969D}{16}+\frac{1855}{32(2D-7)}-\frac{3}{2(D-3)}\right.\nonumber \\
&&\left.\qquad \qquad-\frac{730}{(D-4)}-\frac{720}{(D-4)^2}-\frac{288}{(D-4)^3}-\frac{3079}{32}\right)\nonumber \\
&&- C_{6,2}
\left(\frac{D^2}{16}-\frac{21D}{32}-\frac{53}{64(2D-7)}+\frac{29}{64}\right)
\Biggr]\nonumber \\
+C_FC_A\Biggl[
&&-C_{4,1}
\left(\frac{D^2}{16}-\frac{7D}{32}+\frac{265}{64(2D-7)}+\frac{1}{3(D-1)}\right.\nonumber \\
&&\left.\qquad \qquad+\frac{53}{3(D-4)}+\frac{16}{(D-4)^2}+\frac{367}{64}\right)\nonumber \\
&&- B_{3,1} 
\left(\frac{75D^2}{16}-\frac{1129D}{32}+\frac{1855}{64(2D-7)}+\frac{1}{4(D-3)}\right.\nonumber \\
&&\left.\qquad \qquad-\frac{241}{(D-4)}-\frac{228}{(D-4)^2}-\frac{96}{(D-4)^3}-\frac{903}{64}\right)\nonumber \\
&&+C_{6,2}
\left(\frac{D^2}{32}-\frac{21D}{64}-\frac{53}{128(2D-7)}+\frac{29}{128}\right)
\Biggr]\nonumber \\
+C_FN_F\Biggl[
&&-C_{4,1}
\left(D+\frac{2}{3(D-1)}+\frac{4}{3(D-4)}-2\right)
\Biggr] \, .
\end{eqnarray}
}

As in the one-loop case, the all-order result is obtained by replacing $B_{4,2}$, $B_{3,1}$, $C_{4,1}$ and $C_{6,2}$ with $A_{2,\mathrm{LO}}^2$, $A_3$, $A_4$ and $A_6$ from Ref.~\cite{GHM}, respectively.\\
Inserting the expansion of the two-loop master integrals and keeping terms through to ${\cal O}(\eps^2)$ yields
{\allowdisplaybreaks
\begin{eqnarray}
{\cal F}_2 = C_F^2 \Biggl[
&&\frac{2}{\eps^4}
-\frac{1}{\eps^2}\left(2\zeta_2-4\right)
-\frac{1}{\eps}\left(\frac{64\zeta_3}{3}-6\zeta_2-8\right)-\left(13\zeta_2^2-12\zeta_2+30\zeta_3-22\right)\nonumber \\
&&-\eps\left(\frac{96\zeta_2^2}{5}-\frac{112\zeta_2\zeta_3}{3}-36\zeta_2+\frac{404\zeta_3}{3}+\frac{184\zeta_5}{5}-64\right)\nonumber \\
&&+\eps^2\left(\frac{223\zeta_2^3}{5}-\frac{426\zeta_2^2}{5}+2\zeta_2\zeta_3+\frac{2608\zeta_3^2}{9}+106\zeta_2-\frac{1744\zeta_3}{3}-126\zeta_5+192\right) 
%\nonumber \\
%&&-\eps^3\left(\frac{51\zeta_2^3}{35}-\frac{5488\zeta_2^2\zeta_3}{15}+\frac{1884\zeta_2^2}{5}-100\zeta_2\zeta_3-\frac{184\zeta_2\zeta_5}{5}-482\zeta_3^2\right.\nonumber \\
%&&\left.\qquad \quad-312\zeta_2+\frac{7304\zeta_3}{3}+\frac{2288\zeta_5}{5}-\frac{8942\zeta_7}{7}-584\right) 
\Biggr]\nonumber \\
+C_FC_A\Biggl[
&&-\frac{11}{6\eps^3}
+\frac{1}{\eps^2}\left(\zeta_2-\frac{67}{18}\right)
-\frac{1}{\eps}\left(\frac{11\zeta_2}{6}-13\zeta_3+\frac{220}{27}\right)\nonumber \\
&&+\left(\frac{44\zeta_2^2}{5}-\frac{103\zeta_2}{18}+\frac{305\zeta_3}{9}-\frac{1655}{81}\right)\nonumber \\
&&+\eps\left(\frac{1171\zeta_2^2}{60}-\frac{89\zeta_2\zeta_3}{3}-\frac{490\zeta_2}{27}+\frac{2923\zeta_3}{27}+51\zeta_5-\frac{12706}{243}\right)\nonumber \\
&&-\eps^2\left(\frac{809\zeta_2^3}{70}-\frac{11819\zeta_2^2}{180}+\frac{127\zeta_2\zeta_3}{9}+\frac{569\zeta_3^2}{3}+\frac{4733\zeta_2}{81}\right.\nonumber \\
&&\left.\qquad \quad-\frac{30668\zeta_3}{81}-\frac{2411\zeta_5}{15}+\frac{99770}{729}\right) 
%\nonumber \\
%&&+\eps^3\left(
%\frac{5401\zeta_2^3}{140}
%-\frac{7103\zeta_2^2\zeta_3}{30}
%+\frac{6410\zeta_2^2}{27}
%-\frac{2117\zeta_2\zeta_3}{27}
%-\frac{497\zeta_2\zeta_5}{5}
%-\frac{9149\zeta_3^2}{27}\right.\nonumber \\
%&&\left.\qquad \quad
%-\frac{44782\zeta_2}{243}
%+\frac{342892\zeta_3}{243}
%+\frac{21877\zeta_5}{45}
%-372\zeta_7 
%-\frac{801694}{2187}\right) 
\Biggr]\nonumber \\
+C_FN_F\Biggl[
&&\frac{1}{3\eps^3}
+\frac{5}{9\eps^2}
+\frac{1}{\eps}\left(\frac{\zeta_2}{3}+\frac{46}{27}\right)
+\left(\frac{5\zeta_2}{9}-\frac{26\zeta_3}{9}+\frac{416}{81}\right)\nonumber \\
&&-\eps\left(\frac{41\zeta_2^2}{30}-\frac{46\zeta_2}{27}+\frac{130\zeta_3}{27}-\frac{3748}{243}\right)\nonumber \\
&&-\eps^2\left(\frac{41\zeta_2^2}{18}+\frac{26\zeta_2\zeta_3}{9}-\frac{416\zeta_2}{81}+\frac{1196\zeta_3}{81}+\frac{242\zeta_5}{15}-\frac{33740}{729}\right) 
%\nonumber \\
%&&-\eps^3\left(
%\frac{127\zeta_2^3}{14}
%+\frac{943\zeta_2^2}{135}
%+\frac{130\zeta_2\zeta_3}{27}
%-\frac{338\zeta_3^2}{27}\right.\nonumber \\
%&&\left.\qquad \quad
%-\frac{3748\zeta_2}{243}
%+\frac{10816\zeta_3}{243}
%+\frac{242\zeta_5}{9}
%-\frac{303676}{2187}\right) 
\Biggr] \, .
\end{eqnarray}
}

After renormalization, we find full agreement with Eq.~(3.6) of Ref.~\cite{lazopoulos} through to ${\cal O}(\eps^0)$ and  provide the next two terms in the expansion.

\section{Calculation of the three-loop form factors}
\label{sec:red}
As any multi-loop computation, the calculation of the $Hb\bar{b}$ three-loop form factor can be separated into multiple steps: Initially, one calculates the matrix elements in terms of three-loop integrals. Next, the algebraic reduction of all three-loop integrals appearing in the relevant Feynman diagrams is performed. Eventually, one computes the remaining master integrals. Let us elaborate on these three steps.

In order to determine the three-loop vertex function, we use \textsc{Qgraf}~\cite{qgraf} to generate the 244 Feynman diagrams contributing to the $Hb\bar{b}$ form factor at three loops. Every diagram is then contracted with the projector of Eq.~\eqref{eq:projq} and can be expressed as a linear combination of many scalar three-loop Feynman integrals with up to nine different propagators. The integrands depend on the three loop momenta and on the two on-shell external momenta, leading to twelve different combinations of scalar products involving loop momenta. Hence, we are left with irreducible scalar products in the numerator since they do not cancel against all linear independent combinations of denominators.

Using relations between different integrals based on integration-by-parts ~\cite{chet1} and Lorentz invariance ~\cite{gr}, the large number of integrals can be expressed in terms of a small number of master integrals. These identities yield large linear systems of equations, which are solved in an iterative manner using lexicographic ordering as suggested by the Laporta algorithm~\cite{laporta}.  From the available implementations of the Laporta algorithm~\cite{laporta,air,fire,reduze,reduze2}, we apply the \textsc{C++} package \textsc{Reduze}~\cite{reduze,reduze2} to carry out the reduction. For this purpose, we define so-called auxiliary topologies, each of which is a set of twelve linearly independent propagators. 

After the reduction, we are left with 22 master integrals. All of them were computed analytically in the past~\cite{masterA,masterB,masterC,masterD} 
and are summarized in detail in Ref.~\cite{Gehrmann:2010} so that they will not be reproduced here. 

\section{Three-loop form factors}
\label{sec:ff}
The unrenormalized three-loop form factor can be decomposed into the following color structures:
\begin{eqnarray}
{\cal F}_3/S_R^3 &=&  
C_F^3 ~X_{C_F^3}
+C_F^2C_A  ~X_{C_F^2C_A} 
+ C_FC_A^2  ~X_{C_FC_A^2} 
+ C_F^2 N_F  ~X_{C_F^2 N_F}
\nonumber \\
&&
+C_FC_A N_F ~X_{C_FC_A N_F}
+C_FN_F^2  ~X_{C_FN_F^2} \, .
\label{eq:fq3lbare}
\end{eqnarray}
It should be noted that, in contrast to the quark form factor for a photonic 
coupling~\cite{BCSSS,Gehrmann:2010}, no contribution 
from the Higgs boson coupling to closed quark loops appears 
at three loops in massless QCD. This contribution 
requires a helicity flip on both the internal and external quark lines, and is consequently mass-suppressed. 
After the reduction of the integrals appearing in the Feynman diagrams, the coefficients $X_i$ of the color structures include linear combinations of master integrals. These coefficients are somewhat lengthy and will not be presented here. Inserting the expansion of the three-loop master integrals and keeping terms through to ${\cal O}(\eps^0)$, we find that the unrenormalized 
three-loop coefficients are given by 
{\allowdisplaybreaks
\begin{eqnarray}
{\cal F}_3 = 
 C_F^3 \Biggl[
&&-\frac{4}{3\eps^6}
+\frac{1}{\eps^4}\left(2\zeta_2-4\right)
-\frac{1}{\eps^3}\left(12\zeta_2-\frac{100\zeta_3}{3}+8\right)\nonumber \\
&&+\frac{1}{\eps^2}\left(\frac{213\zeta_2^2}{10}-26\zeta_2+60\zeta_3-28\right)\nonumber \\
&&+\frac{1}{\eps}\left(\frac{126\zeta_2^2}{5}-\frac{214\zeta_2\zeta_3}{3}-94\zeta_2+\frac{784\zeta_3}{3}+\frac{644\zeta_5}{5}-\frac{238}{3}\right)\nonumber \\
&&-\left(
          \frac{9095\zeta_2^3}{252}
          - \frac{887\zeta_2^2}{10}
          - 202\zeta_2\zeta_3
          + \frac{1826\zeta_3^2}{3}
\right. \nonumber \\
&&    \left.    \qquad  
          + \frac{1085\zeta_2}{3}
          - 538\zeta_3
          - 676\zeta_5
          + \frac{385}{3}
\right)
\Biggr]\nonumber \\
+C_F^2C_A \Biggl[
&&\frac{11}{3\eps^5}
-\frac{1}{\eps^4}\left(2\zeta_2-\frac{67}{9}\right)
+\frac{1}{\eps^3}\left(\frac{11\zeta_2}{6}-26\zeta_3+\frac{539}{27}\right)\nonumber \\
&&-\frac{1}{\eps^2}\left(\frac{83\zeta_2^2}{5}-\frac{631\zeta_2}{18}+135\zeta_3-\frac{4507}{81}\right)\nonumber \\
&&-\frac{1}{\eps}\left(\frac{31591\zeta_2^2}{360}-\frac{215\zeta_2\zeta_3}{3}-\frac{10199\zeta_2}{54}+\frac{1721\zeta_3}{3}+142\zeta_5-\frac{38012}{243}\right)\nonumber \\
&&-\left(
          \frac{18619\zeta_2^3}{1260}
          + \frac{305831\zeta_2^2}{1080}
          + \frac{1663\zeta_2\zeta_3}{18}
          - \frac{1616\zeta_3^2}{3}
 \right. \nonumber \\
&& \left.          \qquad
          - \frac{131161\zeta_2}{162}
          + \frac{17273\zeta_3}{9}
          + \frac{27829\zeta_5}{45}
          - \frac{332065}{729}
\right) \Biggr]\nonumber \\
+C_FC_A^2 \Biggl[
&&-\frac{242}{81\eps^4}
+\frac{1}{\eps^3}\left(\frac{88\zeta_2}{27}-\frac{3254}{243}\right)
-\frac{1}{\eps^2}\left(\frac{88\zeta_2^2}{45}+\frac{553\zeta_2}{81}-\frac{1672\zeta_3}{27}+\frac{9707}{243}\right)\nonumber \\
&&+\frac{1}{\eps}\left(\frac{802\zeta_2^2}{15}-\frac{88\zeta_2\zeta_3}{9}-\frac{15983\zeta_2}{243}+\frac{8542\zeta_3}{27}-\frac{136\zeta_5}{3}
-\frac{385325}{4374}\right)\nonumber \\
&& -\left(
          \frac{6152\zeta_2^3}{189}
          - \frac{100597\zeta_2^2}{540}
          + \frac{980\zeta_2\zeta_3}{9}
          + \frac{1136\zeta_3^2}{9}
 \right. \nonumber \\
&& \left.        \qquad
          + \frac{478157\zeta_2}{1458}
          - \frac{306992\zeta_3}{243}
          - \frac{3472\zeta_5}{9}
          + \frac{1870897}{26244}
\right) 
\Biggr]\nonumber \\
+C_F^2N_F \Biggl[
&&-\frac{2}{3\eps^5}-\frac{10}{9\eps^4}-\frac{1}{\eps^3}\left(\frac{\zeta_2}{3}+\frac{104}{27}\right)-\frac{1}{\eps^2}\left(\frac{53\zeta_2}{9}-\frac{146\zeta_3}{9}+\frac{865}{81}\right)\nonumber \\
&&+\frac{1}{\eps}\left(\frac{337\zeta_2^2}{36}-\frac{736\zeta_2}{27}+\frac{1882\zeta_3}{27}-\frac{15511}{486}\right)\nonumber \\
&&+\left(\frac{15769\zeta_2^2}{540}-\frac{343\zeta_2\zeta_3}{9}-\frac{16885\zeta_2}{162}+\frac{27812\zeta_3}{81}+\frac{278\zeta_5}{45}-\frac{307879}{2916}\right) 
\Biggr]\nonumber \\
+C_FC_AN_F \Biggl[
&&\frac{88}{81\eps^4}-\frac{1}{\eps^3}\left(\frac{16\zeta_2}{27}-\frac{1066}{243}\right)+\frac{1}{\eps^2}\left(\frac{316\zeta_2}{81}-\frac{256\zeta_3}{27}+\frac{3410}{243}\right)\nonumber \\
&&-\frac{1}{\eps}\left(\frac{44\zeta_2^2}{5}-\frac{5033\zeta_2}{243}+\frac{5140\zeta_3}{81}-\frac{90305}{2187}\right)\nonumber \\
&&-\left(\frac{3791\zeta_2^2}{135}-\frac{368\zeta_2\zeta_3}{9}-\frac{63571\zeta_2}{729}+\frac{23762\zeta_3}{81}+\frac{208\zeta_5}{3}-\frac{1451329}{13122}\right) 
\Biggr]\nonumber \\
+C_FN_F^2 \Biggl[
&&-\frac{8}{81\eps^4}-\frac{80}{243\eps^3}-\frac{1}{\eps^2}\left(\frac{4\zeta_2}{9}+\frac{32}{27}\right)-\frac{1}{\eps}\left(\frac{40\zeta_2}{27}-\frac{136\zeta_3}{81}+\frac{9616}{2187}\right)\nonumber \\
&&-\left(\frac{83\zeta_2^2}{135}+\frac{16\zeta_2}{3}-\frac{1360\zeta_3}{243}+\frac{109528}{6561}\right) 
\Biggr] \;.
\end{eqnarray}

The UV renormalization of the $Hb\bar{b}$ form factor has been derived in Section~\ref{sec:renorm}. Applying Eq.~\eqref{eq:reng} yields the expansion coefficients of the renormalized form factors. In the time-like kinematics, the real part reads
{\allowdisplaybreaks
\begin{eqnarray}
\operatorname{Re} {F}_3 = 
 C_F^3 \Biggl[
&&-\frac{4}{3\eps^6}
-\frac{6}{\eps^5}
+\frac{1}{\eps^4}\left(38\zeta_2-13\right)
+\frac{1}{\eps^3}\left(66\zeta_2+\frac{100\zeta_3}{3}-23\right)\nonumber \\
&&-\frac{1}{\eps^2}\left(\frac{1947\zeta_2^2}{10}-\frac{191\zeta_2}{2}-124\zeta_3+\frac{235}{4}\right)\nonumber \\
&&+\frac{1}{\eps}\left(\frac{861\zeta_2^2}{5}-\frac{2914\zeta_2\zeta_3}{3}+\frac{899\zeta_2}{4}+\frac{1117\zeta_3}{3}+\frac{644\zeta_5}{5}-\frac{550}{3}\right)\nonumber \\
&&-\left(
            \frac{19301\zeta_2^3}{252}
          - \frac{4495\zeta_2^2}{8}
          + 2298\zeta_2\zeta_3
          + \frac{1826\zeta_3^2}{3}
 \right. \nonumber \\
&&    \left.    \qquad  
          - \frac{3635\zeta_2}{6}
          - \frac{1877\zeta_3}{2}
          - \frac{3932\zeta_5}{5}
          + \frac{1060}{3}
\right)
\Biggr]\nonumber \\
+C_F^2C_A \Biggl[
&& - \frac{11}{\eps^5}
-\frac{1}{\eps^4}\left(
            2 \zeta_2
          + \frac{361}{18}
\right)
+ \frac{1}{\eps^3}\left(
            \frac{181\zeta_2}{2}
          - 26 \zeta_3
          + \frac{79}{54}
\right)\nonumber \\
&&+\frac{1}{\eps^2}\left(
            \frac{187\zeta_2^2}{5}
          - \frac{2789\zeta_2}{18}
          - \frac{158\zeta_3}{9}
          + \frac{4699}{324}
\right)\nonumber \\
&&-\frac{1}{\eps}\left(
            \frac{8267\zeta_2^2}{72}
          - \frac{2321\zeta_2\zeta_3}{3}
          + \frac{28031\zeta_2}{108}
          + \frac{1135\zeta_3}{3}
          + 142\zeta_5
          - \frac{16823}{972}
\right)\nonumber \\
&&+\left(
            \frac{239933\zeta_2^3}{1260}
          + \frac{78529\zeta_2^2}{270}
          + \frac{3917\zeta_2\zeta_3}{2}
          + \frac{1616\zeta_3^2}{3}
 \right. \nonumber \\
&& \left.          \qquad
          - \frac{30463\zeta_2}{81}
          - \frac{7765\zeta_3}{6}
          - \frac{4514\zeta_5}{9}
          + \frac{31618}{729}
\right) \Biggr]\nonumber \\
+C_FC_A^2 \Biggl[
&&-\frac{1331}{81\eps^4}
-\frac{1}{\eps^3}\left(
            \frac{110\zeta_2}{27}
          - \frac{2866}{243}
\right)
-\frac{1}{\eps^2}\left(
            \frac{88\zeta_2^2}{45}
          - \frac{1625\zeta_2}{81}
          + \frac{902\zeta_3}{27}
          - \frac{11669}{486}
\right)\nonumber \\
&&-\frac{1}{\eps}\left(
            \frac{166\zeta_2^2}{15}
          + \frac{88\zeta_2\zeta_3}{9}
          + \frac{7163\zeta_2}{243}
          - \frac{3526\zeta_3}{27}
          + \frac{136\zeta_5}{3}
          + \frac{139345}{8748}
\right)\nonumber \\
&& +\left(
            \frac{19136\zeta_2^3}{945}
          - \frac{3137\zeta_2^2}{135}
          - \frac{1258\zeta_2\zeta_3}{3}
          - \frac{1136\zeta_3^2}{9}
\right. \nonumber \\
&& \left.        \qquad  
          + \frac{380191\zeta_2}{1458}
          + \frac{107648\zeta_3}{243}
          + \frac{106\zeta_5}{9}
          + \frac{5964431}{26244}
\right) 
\Biggr]\nonumber \\
+C_F^2N_F \Biggl[
&& \frac{2}{\eps^5}+\frac{35}{9\eps^4}
-\frac{1}{\eps^3}\left(
          17\zeta_2
          + \frac{23}{27}
\right)+\frac{1}{\eps^2}\left(
            \frac{199\zeta_2}{9}
          - \frac{110\zeta_3}{9}
          - \frac{641}{162}
\right)\nonumber \\
&&+\frac{1}{\eps}\left(
            \frac{577\zeta_2^2}{36}
          + \frac{3235\zeta_2}{54}
          + \frac{442\zeta_3}{27}
          - \frac{967}{486}
\right)\nonumber \\
&&-\left(
            \frac{8822\zeta_2^2}{135}
          + 85\zeta_2\zeta_3
          - \frac{22571\zeta_2}{162}
          - \frac{15131\zeta_3}{81}
          + \frac{386\zeta_5}{9}
          + \frac{145375}{2916}
\right) 
\Biggr]\nonumber \\
+C_FC_AN_F \Biggl[
&& \frac{484}{81\eps^4}+\frac{1}{\eps^3}\left(
            \frac{20\zeta_2}{27}
          - \frac{752}{243}
\right)-\frac{1}{\eps^2}\left(
            \frac{476\zeta_2}{81}
          - \frac{212\zeta_3}{27}
          + \frac{2068}{243}
\right)\nonumber \\
&&+\frac{1}{\eps}\left(
            \frac{44\zeta_2^2}{15}
          + \frac{2594\zeta_2}{243}
          - \frac{964\zeta_3}{81}
          - \frac{8659}{2187}
\right)\nonumber \\
&&-\left(
            \frac{836\zeta_2^2}{135}
          - \frac{148\zeta_2\zeta_3}{3}
          + \frac{59999\zeta_2}{729}
          + \frac{2860\zeta_3}{27}
          + \frac{4\zeta_5}{3}
          + \frac{521975}{13122}
\right) 
\Biggr]\nonumber \\
+C_FN_F^2 \Biggl[
&&-\frac{44}{81\eps^4}-\frac{8}{243\eps^3}+\frac{1}{\eps^2}\left(
            \frac{4}{9}\zeta_2
          + \frac{46}{81}
\right)-\frac{1}{\eps}\left(
            \frac{20}{27}\zeta_2
          + \frac{8}{81}\zeta_3
          - \frac{2417}{2187}
\right)\nonumber \\
&&+\left(
            \frac{172}{135}\zeta_2^2
          + \frac{388}{81}\zeta_2
          - \frac{200}{243}\zeta_3
          + \frac{2072}{6561}
\right) 
\Biggr]\;. \label{eq:f3qr}
\end{eqnarray}}

For the sake of completeness, let us state that the imaginary part of the UV renormalized three-loop form factor is given by
{\allowdisplaybreaks
\begin{eqnarray}
\frac{\operatorname{Im} {F}_3}{\pi} = 
 C_F^3 \Biggl[
&&-\frac{4}{\eps^5}
-\frac{12}{\eps^4}
+\frac{1}{\eps^3}\left(42\zeta_2-21\right)+\frac{1}{\eps^2}\left(24\zeta_2+100\zeta_3-\frac{93}{2}\right)\nonumber \\
&&-\frac{1}{\eps}\left(\frac{873\zeta_2^2}{10}-\frac{15\zeta_2}{2}-308\zeta_3+141\right)\nonumber \\
&&+\left(
            372\zeta_2^2
          - 1114\zeta_2\zeta_3
          - \frac{177\zeta_2}{4}
          + 985\zeta_3
          + \frac{1932\zeta_5}{5}
          - \frac{773}{2}
\right)
\Biggr]\nonumber \\
+C_F^2C_A \Biggl[
&& -\frac{55}{3\eps^4}
- \frac{1}{\eps^3}\left(
            6\zeta_2
          - \frac{1}{3}
\right)
+\frac{1}{\eps^2}\left(
            \frac{283\zeta_2}{6}
          - 78\zeta_3
          + \frac{715}{18}
\right)\nonumber \\
&&+\frac{1}{\eps}\left(
            \frac{21\zeta_2^2}{5}
          - \frac{502\zeta_2}{3}
          - \frac{1531\zeta_3}{9}
          + \frac{1768}{27}
\right)\nonumber \\
&&-\left(
            \frac{5669\zeta_2^2}{40}
          - 917\zeta_2\zeta_3
          - \frac{253\zeta_2}{36}
          + \frac{4222\zeta_3}{3}
          + 426\zeta_5
          - \frac{35539}{162}
\right) \Biggr]\nonumber \\
+C_FC_A^2 \Biggl[
&&-\frac{242}{27\eps^3}
-\frac{1}{\eps^2}\left(
            \frac{44\zeta_2}{9}
          - \frac{2086}{81}
\right)
-\frac{1}{\eps}\left(
            \frac{88\zeta_2^2}{15}
          - \frac{536\zeta_2}{27}
          + \frac{44\zeta_3}{9}
          + \frac{245}{9}
\right)\nonumber \\
&& +\left(
            2\zeta_2^2
          - \frac{88\zeta_2\zeta_3}{3} 
          + \frac{1036\zeta_2}{81}
          + \frac{13900\zeta_3}{27}
          - 136\zeta_5
          - \frac{10289}{1458}
\right) 
\Biggr]\nonumber \\
+C_F^2N_F \Biggl[
&& \frac{10}{3\eps^4}+\frac{2}{3\eps^3}
-\frac{1}{\eps^2}\left(
            \frac{29\zeta_2}{3}
          + \frac{71}{9}
\right)
+\frac{1}{\eps}\left(
            \frac{76\zeta_2}{3}
          - \frac{74\zeta_3}{9}
          - \frac{403}{27}
\right)\nonumber \\
&&+\left(
            \frac{3\zeta_2^2}{4}
          + \frac{487\zeta_2}{18}
          + \frac{1192\zeta_3}{9}
          - \frac{9649}{162}
\right) 
\Biggr]\nonumber \\
+C_FC_AN_F \Biggl[
&& \frac{88}{27\eps^3}+\frac{1}{\eps^2}\left(
            \frac{8\zeta_2}{9}
          - \frac{668}{81}
\right)-\frac{1}{\eps}\left(
            \frac{80\zeta_2}{27}
          - \frac{56\zeta_3}{9}
          - \frac{418}{81}
\right)\nonumber \\
&&+\left(
            \frac{12\zeta_2^2}{5}
          - \frac{196\zeta_2}{81}
          - \frac{724\zeta_3}{9}
          + \frac{7499}{729}
\right) 
\Biggr]\nonumber \\
+C_FN_F^2 \Biggl[
&&-\frac{8}{27\eps^3}+\frac{40}{81\eps^2}+\frac{8}{81\eps}
-\left(
            \frac{16}{27}\zeta_3
          + \frac{928}{729}
\right) 
\Biggr] \, .
\end{eqnarray}}

It should be stressed that for every color structure of Eq.~\eqref{eq:fq3lbare}, the coefficients of the leading poles in $\e$ agree with the ones of the $\gamma^*q\bar q$ form 
factor of Ref.~\cite{Gehrmann:2010}, as expected.

\section{Infrared pole structure}
\label{sec:ir}
A more powerful check consists in analyzing the complete infrared pole structure of our three-loop results. 

 As outlined in Refs.~\cite{catani,sterman,MMV1,Becher:2009cu,Gardi:2009qi}, it can be predicted from infrared factorization properties of QCD. Accordingly, the infrared pole structure of the renormalized $Hb\bar{b}$ form factors $F_1$, $F_2$ and $F_3$ can be derived from the same formulae as for the $\gamma^*q\bar q$ form factor in Ref.~\cite{Gehrmann:2010}. They read
\begin{eqnarray}
\label{polef1q}
\Poles{(F_1)} &=& 
-\frac{C_F \gamma^{\rm cusp}_0}{2 \eps^2}+\frac{\gamma^q_0}{\eps}\,,\\
\label{polef2q}
\Poles{(F_2)} &=& 
\frac{3 C_F \gamma^{\rm cusp}_0 \beta_0}{8 \eps^3}+\frac{1}{\eps^{2}}\biggl(-\frac{\beta_0 \gamma^q_0}{2}-\frac{C_F 
\gamma^{\rm cusp}_1}{8}\biggr)
+\frac{\gamma^q_1}{2 \eps} +\frac{\left(F_1\right)^2}{2}\,,\\
\label{polef3q}
\Poles{(F_3)} &=& 
-\frac{11 \beta_0^2 C_F \gamma^{\rm cusp}_0}{36 \eps^4}+\frac{1}{\eps^{3}}
\biggl( \frac{5 \beta_0 C_F \gamma^{\rm cusp}_1}{36}+\frac{\beta_0^2 \gamma^q_0}{3}+\frac{2 C_F \gamma^{\rm cusp}_0 \beta_1}{9}\biggr)\nonumber \\ & &
+\frac{1}{\eps^{2}}\biggl(-\frac{\beta_0 \gamma^q_1}{3}-\frac{C_F \gamma^{\rm cusp}_2}{18}-\frac{\beta_1 \gamma^q_0}{3}\biggr)
+\frac{\gamma^q_2}{3 \eps}
-\frac{\left(F_1\right)^3}{3}+F_2 F_1\,.
\end{eqnarray}
These equations require knowing the coefficients $\gamma^{\rm cusp}_i$ of the cusp soft anomalous dimension up to three loops~\cite{MMV1}:
\begin{eqnarray}
\gamma^{\rm cusp}_0 &=& 4\,,\\
\gamma^{\rm cusp}_1 &=& 
C_A\bigg(\frac{268}{9}-\frac{4\pi^2}{3}\bigg) 
-\frac{40N_F}{9}\,,\\
\gamma^{\rm cusp}_2 &=& 
C_A^2\bigg(\frac{490}{3}-\frac{536\pi^2}{27}+\frac{44\pi^4}{45}+\frac{88\zeta_3}{3}\bigg) 
+C_AN_F\bigg(-\frac{836}{27}+\frac{80\pi^2}{27}-\frac{112\zeta_3}{3}\bigg)\nonumber \\ & &
+C_FN_F\bigg(-\frac{110}{3}+32\zeta_3\bigg) 
-\frac{16N_F^2}{27}\,.
\end{eqnarray}
Moreover, $\gamma^{q}_i$ denotes the coefficients of the quark collinear anomalous dimension. To three-loop order, they are given by~\cite{Becher:2006mr,Becher:2009qa}:
{\allowdisplaybreaks
\begin{eqnarray}
\gamma^q_0 &=& -3C_F\,,\\
\gamma^q_1 &=& 
C_F^2\biggl(-\frac{3}{2}+2\pi^2-24\zeta_3\bigg) 
+C_FC_A\biggl(-\frac{961}{54}-\frac{11\pi^2}{6}+26\zeta_3\bigg)\nonumber \\ & &
+C_FN_F\biggl(\frac{65}{27}+\frac{\pi^2}{3}\bigg)\,,\\
\gamma^q_2 &=& 
C_F^2N_F\biggl(\frac{2953}{54}-\frac{13\pi^2}{9}-\frac{14\pi^4}{27}+\frac{256\zeta_3}{9}\bigg) 
+C_FN_F^2\biggl(\frac{2417}{729}-\frac{10\pi^2}{27}-\frac{8\zeta_3}{27}\bigg)\nonumber \\ & &
+C_FC_AN_F\biggl(-\frac{8659}{729}+\frac{1297\pi^2}{243}+\frac{11\pi^4}{45}-\frac{964\zeta_3}{27}\bigg) \nonumber \\ & &
+C_F^3\biggl(-\frac{29}{2}-3\pi^2-\frac{8\pi^4}{5}-68\zeta_3+\frac{16\pi^2\zeta_3}{3}+240\zeta_5\bigg)\nonumber \\ & &
+C_AC_F^2\biggl(-\frac{151}{4}+\frac{205\pi^2}{9}+\frac{247\pi^4}{135}-\frac{844\zeta_3}{3}-\frac{8\pi^2\zeta_3}{3}-120\zeta_5\bigg)\nonumber \\ & &
+C_A^2C_F\biggl(-\frac{139345}{2916}-\frac{7163\pi^2}{486}-\frac{83\pi^4}{90}+\frac{3526\zeta_3}{9}-\frac{44\pi^2\zeta_3}{9}-136\zeta_5\bigg)\;.
\end{eqnarray}
}

The deepest infrared pole for the $i$-loop form factor $F_i$ is proportional to $\eps^{-2i}$. Due to the last term of Eq.~\eqref{polef3q}, we thus need to include the renormalized form factors $F_1$ through to ${\cal O}(\eps^3)$ and $F_2$ through to ${\cal O}(\eps)$, both stated in 
Section \ref{sec:renorm} above. 
 In doing so, we succeed in reproducing the infrared poles of the 
 renormalized form factor up to three loops. 

\section{Conclusions}
\label{sec:conc}
In this paper, we have derived the three-loop QCD corrections to the form factor describing the Yukawa 
coupling of a Higgs boson to a pair of bottom quarks. We neglect the bottom quark mass in internal 
propagators and external states, which is justified by the large mass hierarchy between the Higgs boson  and the bottom quark. The pole structure of our result is in agreement with the prediction of 
infrared factorization formulae~\cite{catani,sterman,MMV1,Becher:2009cu,Gardi:2009qi}.

Our results can be applied to derive the third-order QCD corrections 
to Higgs boson production from bottom quark fusion and to the fully differential description of
Higgs boson decays into bottom quarks. Besides the three-loop corrections derived here, these 
reactions also require two-loop corrections to the matrix element for $Hb\bar bg$, derived recently 
in Ref.~\cite{Ahmed:2014} and higher multiplicity tree-level 
and one-loop matrix elements that can by now be 
derived using standard methods. The integration of all subprocess contributions over the relevant 
phase spaces is far from trivial, and methods are currently under intensive 
development~\cite{n3lomethods}.

 A more imminent application is 
the N$^3$LO soft-virtual threshold approximation to Higgs boson production in 
bottom quark fusion, using the recently derived result for Higgs boson production in 
gluon fusion to this 
order~\cite{claudehiggs}, combined with universal factorization 
properties~\cite{ravindran1,grazziniRV}.

\section*{Acknowledgements}
We would like to thank V.\ Ravindran for useful discussions leading to this project and M.\ Wiesemann for interesting comments throughout. 
This research was supported in part by
the Swiss National Science Foundation (SNF) under contract
200020-149517, as well as  by the European Commission through the 
``LHCPhenoNet" Initial Training Network PITN-GA-2010-264564 and the ERC Advanced Grant
``MC@NNLO" (340983).

\end{document}